\documentclass[10pt,draft]{dis03}
\usepackage{epsf,amsmath}

\newcommand{\be}{\begin{equation}}

\newcommand{\ee}{\end{equation}}

\newcommand{\bea}{\begin{eqnarray}}

\newcommand{\eea}{\end{eqnarray}}

\newcommand{\p}{\vec{p}}

\begin{document}

\title{DIFFRACTIVE ELECTROPRODUCTION OF TWO MESONS\\ 
SEPARATED BY A LARGE RAPIDITY GAP\footnote{Presented by L.~Szymanowski
at the DIS 03 conference
}}

\author{D.Yu.~IVANOV${}^{1,2}$,
 B.~PIRE${}^{3}$,
 L.~SZYMANOWSKI${}^{4}$ \\
 and
 O.V.~TERYAEV ${}^{5}$\\
$¥^1$
   I T P, Universit{\"a}t Regensburg,  D-93040 Regensburg, Germany \\
$ ^2$ 
Institute of Mathematics, 630090 Novosibirsk, Russia \\                    
 $^3$ 
CPhT, {\'E}cole Polytechnique, F-91128 Palaiseau, France\footnote{
  Unit{\'e} mixte C7644 du CNRS.}\\                  
$^4$ 
 So{\l}tan Institute for Nuclear Studies,
Ho\.za 69, 00-681 Warsaw, Poland \\                    
$^5$ 
Bogoliubov Lab. of Theoretical Physics, JINR, 141980 Dubna, Russia\\
  }

\maketitle
\vskip.1in
\begin{abstract}
\noindent We consider the process
$ \gamma ^* N \to M_1 M_2 N'$ with a large rapidity gap between the two  mesons 
$M_1$ and $ M_2$¥. Within the QCD collinear approximation, 
the  scattering amplitude may be written as a convolution 
of an impact factor describing the $\gamma ^* \to M_1$ transition
and an amplitude describing the $N\to M_2 N'$ collinear process. 
\end{abstract}

\section{The Process} 
Let us consider \cite{Ivanov:2002jj} the  process
$¥A N \to M_{1}¥ M_{2}¥ N'$
shown in Fig. 1 of scattering
of a particle $A$, e.g.  being a virtual or real photon, on a
nucleon $N$, which leads via two gluon exchange to the production
of  particle $M_{1}¥$ ( meson or pair of mesons) separated by a large
rapidity gap from another produced meson $M_{2}¥$ and the scattered nucleon $N'$.
We consider the kinematical region where the rapidity gap between $M_{2}$ and
$N'$ is much smaller than the one between
$M_{1}¥$ and $M_{2}$, that is the energy of the system ($M_{2}\; -\; N'$) is smaller
than the energy of the system ($M_{1}\; -\; M_{2}$) but still large enough to
justify our approach  (in particular much larger than baryonic resonance masses). 

Such a process is a representative of a new class of hard
reactions whose QCD description  involves at the same time
the impact factor appearing naturally
in Regge-type perturbative description 
based on the BFKL evolution \cite{BFKL} and  the collinear
distributions which have been introduced for describing deeply
 virtual Compton scattering \cite{GPD} and whose evolutions are governed
by DGLAP-ERBL
equations.
%
%%%%%%%%%%%%%%%%%%     FIGURE 1          %%%%%%%%%%%%%%%%%%%%%%%%%%%%
\begin{figure}[t]
\centerline{\epsfxsize5.0cm\epsffile{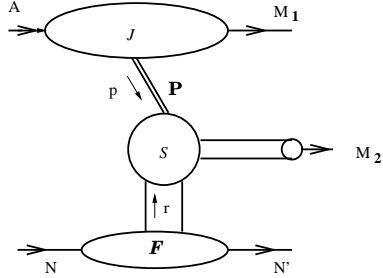}}
\caption[]{\small
Factorization of the process $A\;N \to \;M_{1}\;M_{2}\;N'$ in the asymmetric
kinematics
discussed in the text. ${\bf P}$ is the hard Pomeron.
 }
\label{fig:1}
\end{figure}
%%%%%%%%%%%%%%%%%%%%%%%%%%%%%%%%%%%%%%%%%%%%%%%%%%%%%%%%%%%%%%%%%%%%%%

In order to make our discussion  clear  let us 
consider  a reference
process with all longitudinally polarized vector particles
\be
\label{2mesongen}
\gamma^*_L (q)\;\; N (p_2) \to  \rho_L^0(q_\rho)\;\; \rho_L^+(p_\rho)
N'(p_{2'})\;,
\ee
which involves the emission of two gluons
 in the $\gamma^*_L \to \rho_L$ transition and 
which is more founded theoretically from the point of view 
of the collinear factorization.
We choose a charged vector meson $\rho^+$ to select quark antiquark exchange
with the nucleon.

\section{Kinematics}
Let us  summarize  the kinematics of the process
(\ref{2mesongen}). 
We introduce two light-like Sudakov vectors
$p_{1/2}$. The momenta are parametrized as follows~:
\bea
&& q^\mu = p_1^\mu -
\frac{Q^2}{s}p_2^\mu\;,\;\;\;\;q^2=-Q^2\,,\;\;\;\;s=2(p_1p_2)\;,\nonumber
\\
&& q_\rho^\mu = \alpha p_1^\mu +
\frac{\p^{\;2}}{\alpha s}p_2^\mu + p_\perp^\mu \;,\;\;\;\;
\;p_\perp^2 = -\p^{\;2} \;, 
\nonumber \\
&& p_\rho^\mu = \bar \alpha p_1^\mu + \frac{\p^{\;2}}{\bar \alpha s}p_2^\mu
- p_\perp^\mu \;,\;\;\;\;\;\bar \alpha \equiv 1-\alpha \;, \nonumber \\
&& p_{2'}^\mu = p_2^\mu (1- \zeta)\;,
\label{Sud}
\eea
where $\zeta$ is the skewedness parameter  which can be written as 
\be
\zeta = \frac{1}{s}\,\left(Q^2 + s_1  \right)\,,
\label{zeta}
\ee
and where $¥s_{1}$ is
the two meson invariant mass 
$¥s_1 = (q_\rho + p_\rho)^2 = {\p^{\;2}}/{\alpha \bar \alpha}$.

The $\rho^+(p_\rho)-$meson - target invariant mass equals
$s_2 = (p_\rho + p_{2'})^2 = s\, \bar \alpha\,\left(1-\zeta  \right)\,.
$
The kinematical limit with a large rapidity gap between the two mesons in the
final state is obtained by demanding that $s_1$ is very large
\be
s_1 =s\, \zeta\,,\;\;\;\;s_1 \gg Q^2,\,\,\p^{\;2}\,, 
\label{s1gap}
\ee 
whereas $s_2$ is kept constant but  large enough to justify the use 
of  perturbation theory in the
collinear subprocess ${\cal P} N \to \rho^+_L N'$ and the application of
the GPD framework \cite{GPD}.
In terms of the longitudinal fraction $\alpha$ the kinematics 
with a large rapidity gap corresponds
to taking the limits
$¥\alpha \to 1\,,\;\;\;\bar \alpha s_1 \to \p^{\;2}\,,\;\;\;\;\zeta
\sim
1\,.$
We consider the case when the scattered nucleon 
 gets no transverse momentum in
the process, but  one may allow a finite
momentum transfer, small with respect to $|\p|$, with slight modifications of the formulae.

\vskip.1in
Let us stress that the role of the  hard scale in the
process under discussion  is played by the virtuality $p^2=-\p^2$,
or by the large momentum transfer
in  the two-gluon exchange channel. If additionally the incoming photon 
has non zero, sufficiently large  virtuality $Q^2$,  then the
theoretical description of the processes simplifies even more,
as we can neglect within our approximation 
the contribution of the hadronic component of the photon.

\section{Amplitude}
We have shown that the scattering amplitude ${\cal M}$ of the process
(\ref{2mesongen}) may be calculated in the collinear factorization
approach.
The final result is represented as a convolution (an 
integral over  the longitudinal momentum fractions of the quarks)  of 
the  two amplitudes: the first one describing
the transition $A \to M_{1}$ via two gluon exchange and 
the second one  describing the subprocess
${\cal P}\;N\;\to \;M_{2}\;N'$ which is
closely related to the electroproduction process $\gamma^*\,N \to M_{2}\,N'$
where  collinear factorization
theorems \cite{Collins} allow to separate  the long distance dynamics
expressed through 
GPDs from a perturbatively calculable coefficient function. The hard scale
appearing in the process 
${\cal P}\;N\;\to \;M_{2}\;N'$
 is supplied by the
relatively large  momentum transfer
$p^2$ in the two gluon channel, i.e. by the virtuality of the Pomeron
${\cal P}$.

The scattering amplitude reads :
\be
\label{fact}
{\cal M} = \sum\limits_{p=q,\bar q}\int\limits_0^1 dz\,\int\limits_0^1
du\,\int\limits_0^1 dx_1
T^p_H(x_1,u,z)\,F^p_\zeta(x_1) \phi_{\rho^+}(u)
\phi_{\rho^0}(z)\;.
\ee
Here $F^p_\zeta(x_1)$ is the generalized  parton $p$ distribution
in the
target at zero momentum transfer; $x_1$ and $x_2 = x_1 -\zeta$ are  the momentum
fractions of the emitted and absorbed partons (quarks) of the target,
respectively (as usual the case $x_2 < 0$ is interpreted as an emitted
antiquark).
$\phi_{\rho^+}(u)$ and $\phi_{\rho^0}(z)$ are the distribution
amplitudes of the $\rho^+-$meson and
$\rho^0-$meson, respectively.
$T^p_H(x_1,u,z)$ is
the hard scattering amplitude (the coefficient function).
For clarity of notation we omit in Eq.~(\ref{fact}) the factorization scale
dependence of $T^p_H$, $F^p_\zeta$, $\phi_{\rho^0}$ and $\phi_{\rho^+}$.

Eq.~(\ref{fact}) describes the amplitude in the leading twist. 
  Within this approximation one neglects (in the physical gauge)
the contributions of the higher Fock states in the meson wave functions
and the many parton correlations (higher twist GPD's) in the proton.
Moreover, one can neglect in the hard scattering amplitude the relative
(with respect to a meson momentum) transverse momenta of
partons
%constituent
%quarks 
 (the collinear approximation). This results in the appearence in
the factorization formula (\ref{fact}) of the distribution amplitudes\footnote{or a 
generalized distribution amplitude \cite{GDA} if $¥M_{1}$¥is a pair of
mesons },
i.e. the 
%usual 
light-cone wave functions depending on the relative transverse
momenta of constituents integrated over these momenta up to the collinear
factorization scale.

In the Born approximation the scattering amplitude 
$T^q_H(x_1,u,z)$ for the quark $q$ target
is described by six diagrams. They are calculated for the
on-mass-shell quarks carrying the collinear momenta $x_{1,\,2}p_2$. 
 The on-mass-shell quark (resp. antiquark) entering the $\rho-$mesons
distribution amplitudes $ \phi_{\rho^+}(u)$ and $\phi_{\rho^0}(z)$ 
carry fractions $u$ (resp. $1-u$) and $z$ (resp. 1-z) of the momentum of a
corresponding outgoing
meson, $q_\rho$ and
 $p_\rho$, respectively. 
Moreover, we shown that the scattering
 amplitude 
$T^q_H(x_1,u,z)$ turns out  to be proportional to $ J^{\gamma^*_L \to
\rho_L^0}(u\vec p,\bar u \vec p)$,
i.e. to  the impact
factor for $\gamma^*_L \to \rho_L^0$ transition via the two gluon
(Pomeron) exchange.

We found that the integrals over $x_1$, $u$ and $z$ in the amplitude 
(\ref{fact})
are convergent which justifies 
the validity of the factorization formula.
  Gauge invariance plays here a crucial role by guaranteeing that the  
impact  factor vanishes when $u,\bar
u \to 0$.

All steps of the derivation can be 
immediately applied to the
description of a whole family of processes, in particular 
those involving the chiral-odd GPD \cite{tra},
 e.g. for
\be
\label{2mesontr}
\gamma^*_L (q)\;\; N (p_2) \to  \rho_L^0(q_\rho)\;\; \rho_T^+(p_\rho)
N'(p_{2'})\;,
\ee
which has been the main motivation for the present studies.
The scattering amplitude for the process (\ref{2mesontr}) has the same
general structure  as  for the reference process (\ref{2mesongen});
it involves
 the $\rho_T^+-$meson distribution
amplitude and the generalized transversity distribution in the target.

\section*{Acknowledgements} This work has been supported in part by
the french-polish collaboration agreement Polonium. Work of D.I. was
supported by german DFG and BMBF (06OR984) and in part by 
RFBR 03-02-17734.

\end{document}